\documentclass[%
 reprint,
 amsmath,amssymb,
prl,
]{revtex4-2}

\usepackage{graphicx}
\usepackage{dcolumn}
\usepackage{bm}
\usepackage{hyperref}

\usepackage[percent]{overpic}

\usepackage{tikz}
\usetikzlibrary{math}

\usepackage{pdfpages}
\usepackage{etoolbox} 
\makeatletter
\patchcmd{\@outputpage@head}{\@ifx{\LS@rot\@undefined}{}{\LS@rot}}{}{}{}
\makeatother

\renewcommand{\Re}{\operatorname{Re}}

\newcommand{\diag}{\operatorname{diag}}

\begin{document}


\title{Extended Anderson Criticality in Heavy-Tailed Neural Networks}

\author{Asem Wardak}
\author{Pulin Gong}%
 \email{pulin.gong@sydney.edu.au}
\affiliation{%
 School of Physics, University of Sydney, New South Wales 2006, Australia
}%

\date{\today}

\begin{abstract}
We investigate the emergence of complex dynamics in networks with heavy-tailed connectivity by developing a non-Hermitian random matrix theory.
We uncover the existence of an extended critical regime of spatially multifractal fluctuations between the quiescent and active phases.
This multifractal critical phase combines features of localization and delocalization and differs from the edge of chaos in classical networks by the appearance of universal hallmarks of Anderson criticality over an extended region in phase space.
We show that the rich nonlinear response properties of the extended critical regime can account for a variety of neural dynamics such as the diversity of timescales, providing a computational advantage for persistent classification in a reservoir setting.

\end{abstract}
\maketitle

\paragraph*{Introduction.---}
In a diverse range of physical, biological, financial and ecological systems, complex dynamics fluctuating across multiple scales emerge from a large number of interacting, nonlinear units with heterogeneous properties.
Understanding the organising principles and behavior of such complex dynamics is a longstanding topic of interest across these diverse fields \cite{RevModPhys.80.1275,*Stanley1988,*Ghashghaie1996}.
In neuroscience and machine learning, neural networks with many interacting neurons likewise exhibit complex dynamics with large fluctuations which are critical for their information processing abilities on real-world inputs \cite{Chialvo2010,*NIPS2016_14851003}.
However, the network mechanisms and fundamental computational capabilities of complex neural dynamics remain elusive.

The classical formulation of complex dynamics in systems with many interacting elements is based on randomly connected networks of coupled, nonlinear units with homogeneous connectivity \cite{PhysRevLett.61.259,*PhysRevX.5.041030}.
By employing mathematical approaches such as mean-field theory and random matrix theory, such networks have robustly predicted a phase of chaotic activity with global, homogeneous (i.e.\ delocalized) fluctuations existing adjacent to an ordered, silent regime, enabling the analysis of a wide range of systems with a characteristic scale 
\cite{PhysRevLett.114.088101,*PhysRevE.90.062710,*PhysRevE.82.011903,*PhysRevLett.110.118101}.
The edge of the ordered and chaotic phases gives rise to a range of critical phenomena which is thought to be necessary for these systems to perform useful computations \cite{10.1162/089976604323057443}.
However, growing evidence has shown that coupling heterogeneity is widespread in complex systems such as biological \cite{Shih2020,*10.7554/eLife.57443,*Buzsaki2014} and artificial neural networks \cite{Perez-Nieves2021,Martin2021}, underscoring the need to understand the fundamental dynamical and computational mechanisms of such heterogeneity.

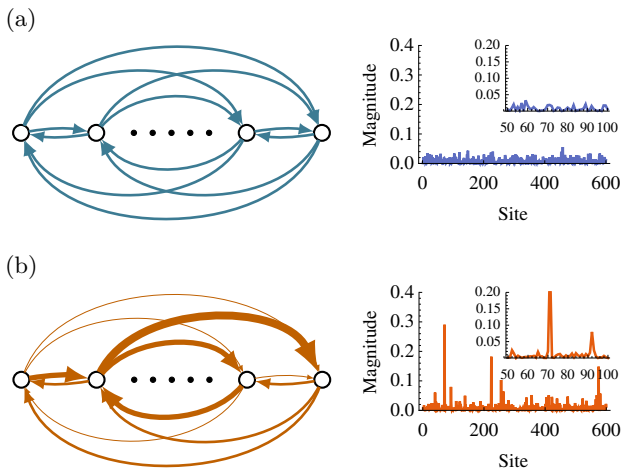
\begin{figure}
    \centering
    \newcommand{\net}[1]
    {
    \def\xs{-2,-1,1,2}
    \def\thicknesses{#1}
    \foreach \x in {-.5,-.25,...,.5} {
        \filldraw[black] (\x,0) circle (1pt);
    }
    \foreach[count=\ix] \x in \xs {
        \draw[thick,black] (\x,0) circle (3pt);
        \foreach[count=\iy] \y in \xs {
            \ifnum \x=\y
            \else
            \pgfmathsetmacro{\xgeqy}{ifthenelse(\x>\y,-1,1)}
            \pgfmathsetmacro{\out}{\xgeqy*90-80/abs(\y-\x)}
            \pgfmathsetmacro{\in}{\xgeqy*90+80/abs(\y-\x)}
            \pgfmathsetmacro{\thickness}{int(\thicknesses[\ix-1][\iy-1])}
            \draw[-latex,shorten >=3pt,shorten <=3pt,line width=\thickness] (\x,0) to[out=\out,in=\in]
            (\y,0);
            \fi
        }
    }
    }
    \begin{tikzpicture}[black!50!cyan]
    \node[black] at (-2,1.5) {(a)};
    \net{
    {{1,1,1,1},
    {1,1,1,1},
    {1,1,1,1},
    {1,1,1,1}}
    }
    \end{tikzpicture}
    \includegraphics[scale=.6]{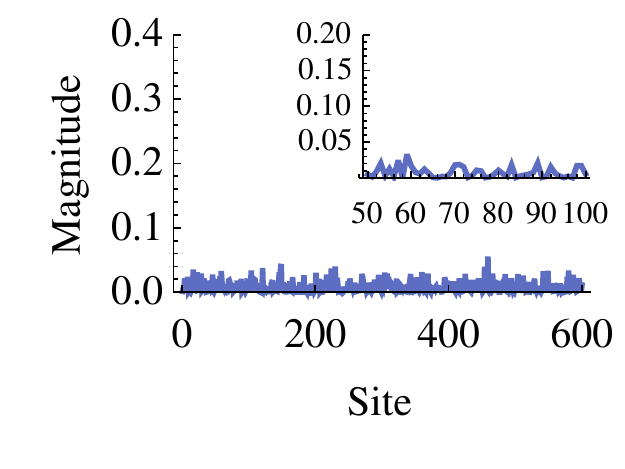}
    \begin{tikzpicture}[black!25!orange]
    \node[black] at (-2,1.5) {(b)};
    \net{
    {{1,2,.5,.5},
    {1.5,1,2,3},
    {.5,2,1,.4},
    {1,1,1,1}}
    }
    \end{tikzpicture}
    \includegraphics[scale=.6]{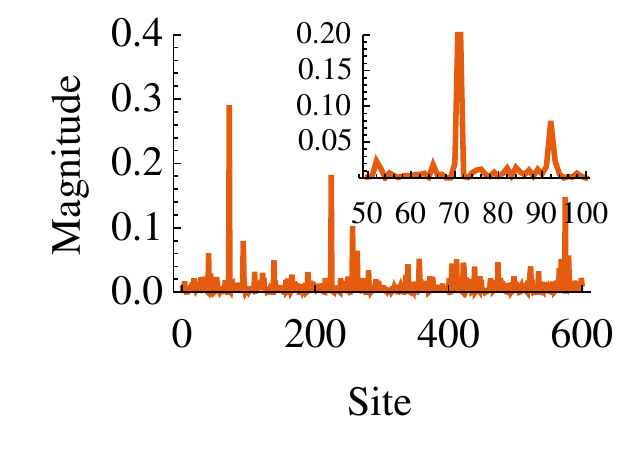}
    \caption{Schematic of the localization of activity fluctuations arising in neural networks.
    (a) In the homogeneous neural network (blue), activity fluctuations are delocalized and spread evenly over the spatial extent of the system ($D_q=1$).
    (b) In the heavy-tailed neural network (red) with heterogeneous weights (line thickness), activity fluctuations are multifractal with a nontrivial distribution of localization over system sites ($0\leq D_q\leq1$) visible over multiple scales (inset).
    }
    \label{fig:schematic}
\end{figure}

Here we study the dynamics of random neural networks with heterogeneous, heavy-tailed connectivity.
After describing the fixed points of the system using a L\'evy mean-field approach, we develop a novel non-Hermitian random matrix theory for column-structured heavy-tailed matrices to analyze the statistical fluctuations of random neural networks around the fixed point.
This theory reveals a new regime with correlated multifractal modes which are neither localized nor delocalized, but have aspects of both (Fig.\ \ref{fig:schematic}).
Multifractality is characterized by the appearance of differing, nontrivial structures appearing simultaneously over a wide variety of scales (first proposed to explain fluctuations in turbulence by Parisi \cite{Benzi_1984}), and is a hallmark of Anderson transitions (criticality) \cite{RevModPhys.80.1355}.
Anderson transitions were first described in the context of disordered electronic systems with localized and metallic (i.e.\ delocalized) phases \cite{PhysRev.109.1492}, and have since been analysed in a broad sense in a wide range of systems including treelike Bethe lattices \cite{Abou_Chacra_1973,*Parisi_2019} and those exhibiting conventional second-order phase transitions \cite{RevModPhys.80.1355}.

We illustrate that the heavy-tailed heterogeneity in connectivity enables Anderson criticality to emerge in a broad parameter regime.
The correlated multifractal modes characteristic of this extended critical regime are able to explain a range of realistic neural dynamics, including correlated fluctuations with low-dimensional features
\cite{MASTROGIUSEPPE2018609,*HUANG2019337},
long-range correlations \cite{HE2014480} and a diversity of timescales \cite{Bernacchia2011}.
Importantly, these correlated multifractal modes provide a profound computational advantage in the setting of real-time reservoir computing by allowing for a persistent form of dimensionality expansion, which is not possible in classical homogeneous systems \cite{PhysRevX.11.021064}.

\paragraph*{Network model and fixed points.---}
\label{sec:network}

We begin by extending the seminal random neural network model with interacting nonlinear units analysed by Sompolinsky et al.\ and others
\cite{PhysRevX.5.041030,PhysRevLett.110.118101,PhysRevLett.114.088101,PhysRevE.90.062710,PhysRevE.82.011903},
which has the dynamics
\begin{align}\label{eq:network}
    h_i'(t) = -h_i(t) + g \sum_{j=1}^N J_{ij} \phi(h_j(t)) ~,
\end{align}
where $h_i(t)$ is the input of the $i$-th neuron at time $t$, $J_{ij}$ is the strength of the connection from neuron $j$ to neuron $i$, and $\phi$ is a scalar nonlinearity that determines the neural firing rate given the input.
The theoretical results in this Letter apply for a wide range of $\phi$ explained below, while the figures use $\phi=\tanh$ for comparability with previous models \cite{PhysRevLett.61.259}.
The coefficient $g$ is the gain parameter of the synaptic input.
Our aim is to investigate networks with heavy-tailed heterogeneity which do not fall within the purview of the Gaussian large-size limit.
Finite-size heterogeneous networks can be investigated to some degree using perturbative finite-order corrections to the homogeneous Gaussian limit \cite{Helias2020,*PhysRevLett.127.158302}.
We instead take an approach which is exact in the large network limit by regarding each $J_{ij}$ as an independent random variable whose second moment is not finite,
so that its probability density has a power-law asymptotic tail,
\begin{align}
    p_{J_{ij}}(x) \stackrel{|x|\to\infty}{\sim} \frac{C_\alpha}{2N|x|^{1+\alpha}} ~,
\end{align}
where $C_\alpha:=\Gamma(1+\alpha)\sin(\pi\alpha/2)/\pi$ and $1<\alpha<2$.
Such heavy-tailed connectivity has been observed in Drosophila \cite{Shih2020,10.7554/eLife.57443}, in successfully trained artificial neural networks \cite{Martin2021} and in spin-glass systems with strong disorder \cite{PhysRevLett.116.010601};
theoretically elucidating the dynamical impact of such heterogeneity has drawn increasing attention \cite{PhysRevLett.125.028101,*PhysRevResearch.3.013083}.
Our approach includes asymmetric L\'evy matrices $J_{ij}\sim L(\alpha,0,0,1/2N)$ with independent $\alpha$-stable \cite{doi:10.1063/1.1318734} elements of scale parameter $1/2N$.
To prevent self-interference we set $J_{ii}=0$, which does not change the overall statistical properties of the connectivity matrix as $N\to\infty$.
In the limit $\alpha\to2$, we recover the classical rate-based random neural network \cite{PhysRevLett.61.259} with independent elements of variance $1/N$.

Applying the generalized central limit theorem to Eq.\ (\ref{eq:network}) in the large network limit yields a L\'evy mean-field theory for the dynamics of the system at long times, whereby each neuron behaves as
\begin{align}\label{eq:MFT}
    h_i'(t) = -h_i(t) + \eta_i(t) ~,
\end{align}
where
$ 
    \eta_i(t)\sim L(\alpha,0,0,
    g^\alpha \langle |\phi(h_j(t))|^\alpha \rangle_j/2 )
$ 
is a time-dependent $\alpha$-stable field generated by randomly weighted inputs from the other neurons.
Setting the left hand side of Eq.\ (\ref{eq:MFT}) to zero gives the fixed points of the network, whose components are distributed as an $\alpha$-stable random variable for all values of the gain $g$.
As we discover in our Jacobian analysis below, the zero fixed point is unstable so that network activity is nonzero and has a macroscopic number of nonzero fixed points at any gain $g$ for $\alpha<2$, in contrast to homogeneous networks ($\alpha=2$) for which the network exhibits a single stable fixed point when $g<1$.
Moreover, since the synaptic input $\eta_i(t)$ has power-law fluctuations over neurons $i$ for heavy-tailed networks with $\alpha<2$, the synaptic input autocorrelation vector $\langle \eta_i(t)\eta_i(t+\tau)\rangle_t$ has the asymptotic tail of an $\alpha/2$-stable distribution over neurons $i$, making the averaged synaptic input autocorrelation $\langle \eta_i(t)\eta_i(t+\tau)\rangle_{i,t}$ infinite in general along with the local-field autocorrelation $\langle h_i(t)h_i(t+\tau)\rangle_{i,t}$.
This non-self-averaging property of heavy-tailed network input differs greatly from classical homogeneous networks where a Gaussian dynamical mean-field theory predicts that all neurons of a sufficiently large population have the same autocorrelation function, and thus the same relaxation timescale \cite{PhysRevX.5.041030}.
On the other hand, the heavy-tailed distribution of local-field autocorrelations over neurons in heterogeneous neural networks allows for a wide reservoir of timescales across neurons of the same population even in the large network limit.

\paragraph{Network stability and heavy-tailed random matrix theory.---}

To determine the local stability of the network around the fixed points predicted by the above mean-field approach, we analyze the Jacobian matrix
$-I+gJ\diag_j(\phi'(h_j))$ obtained from Eq.\ (\ref{eq:network}) where $\diag_j \chi_j$ denotes the diagonal matrix with entries $\chi_j$.
Shifting this Jacobian matrix yields a stability matrix obtained by scaling the columns of $gJ$ by $\phi'(h_j)$, which has the form of a column-structured non-Hermitian random matrix \footnote{For clarity, we refer to $gJ\diag_j(\phi'(h_j))$ as the Jacobian matrix in the remainder of the Letter as its eigenvectors remain unchanged and its eigenvalues are shifted one unit to the right}.
Since the heavy-tailed matrix $J$ (and thus $J\diag_j\chi_j$) has a locally treelike structure \cite{Bordenave2011}, we develop a new cavity approach for column-structured non-Hermitian heavy-tailed random matrices to obtain the spectral density and eigenvector localization properties of the Jacobian \footnote{See Supplemental Material.}.
In recent years, cavity approaches have been used in asymmetrically disordered contexts involving non-Hermitian random ensembles by mapping the problem back to a symmetric, Hermitian system of twice the dimensionality \cite{Lucas_Metz_2019}.
Our cavity approach to the column-structured matrix $J\diag_j\chi_j$ for any $\chi_j$ with $\langle|\chi_j|^\alpha\rangle_j<\infty$ yields the spectral density
\begin{align}\label{eq:RMT_eigenvalues}
    \rho(z) = \frac{y_*^2 - 2|z|^2 y_* \partial_{|z|^2}y_*}{\pi} \left\langle \frac{|\chi_i|^2 SS'}{(|z|^2 + |\chi_i|^2 y_*^2 SS')^2} \right\rangle_i
\end{align}
and inverse participation ratio (see Ref.\ \cite{Note2} for mathematical derivations) describing the spatial localization of the left and right eigenvectors as a function of eigenvalue modulus $|z|$.
$\langle..\rangle_i$ denotes averaging over $i$ and any relevant random variables, while $S,S'\sim L(\alpha/2,1,0,C_\alpha/4C_{\alpha/2})$ are independent, skewed $\alpha/2$-stable random samples, and $y_*$ is found by solving the equation
\begin{align}
    1 = \left\langle\left(\frac{|\chi_i|^2 S}{|z|^2 + y_*^2 |\chi_i|^2 S S'}\right)^{\alpha/2}\right\rangle_i ~.
\end{align}
Our random matrix theory thus unifies both classical results in column-structured non-Hermitian random matrix theory
\cite{PhysRevLett.114.088101,PhysRevLett.97.188104,*PhysRevE.85.066116}
for which $\alpha=2$ and the random variables $S,S'$ reduce to the constant $1$, and unstructured heavy-tailed non-Hermitian random matrices \cite{Bordenave2011} for which $\chi=1$.

\begin{figure*}
    \centering
    \begin{tabular}[b]{ccc}
        \begin{tabular}[b]{c}
            \begin{overpic}[width=.5\columnwidth]{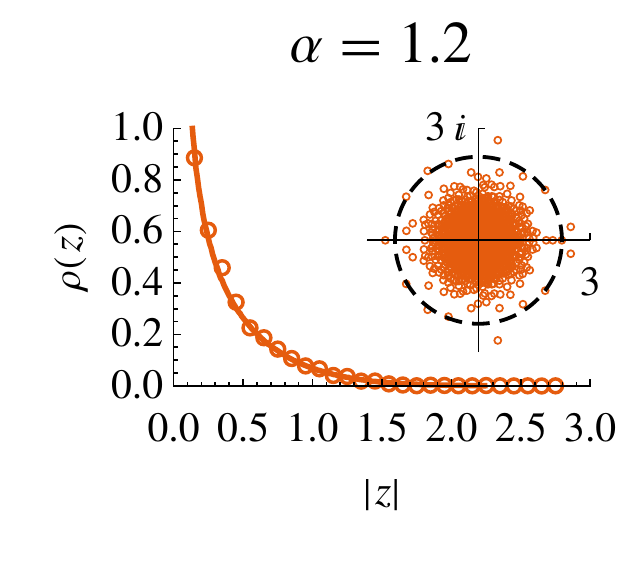}
            \put (0,80) {(a)}
            \end{overpic}
            \\
            \begin{overpic}[width=.5\columnwidth]{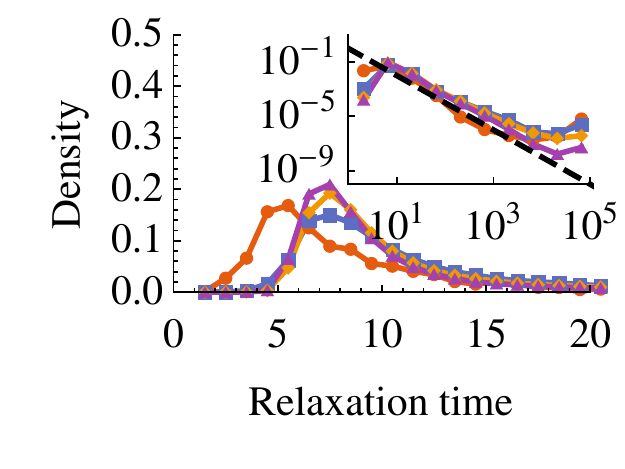}
            \put (0,80) {(b)}
            \end{overpic}
        \end{tabular} &
        \begin{overpic}[width=.75\columnwidth]{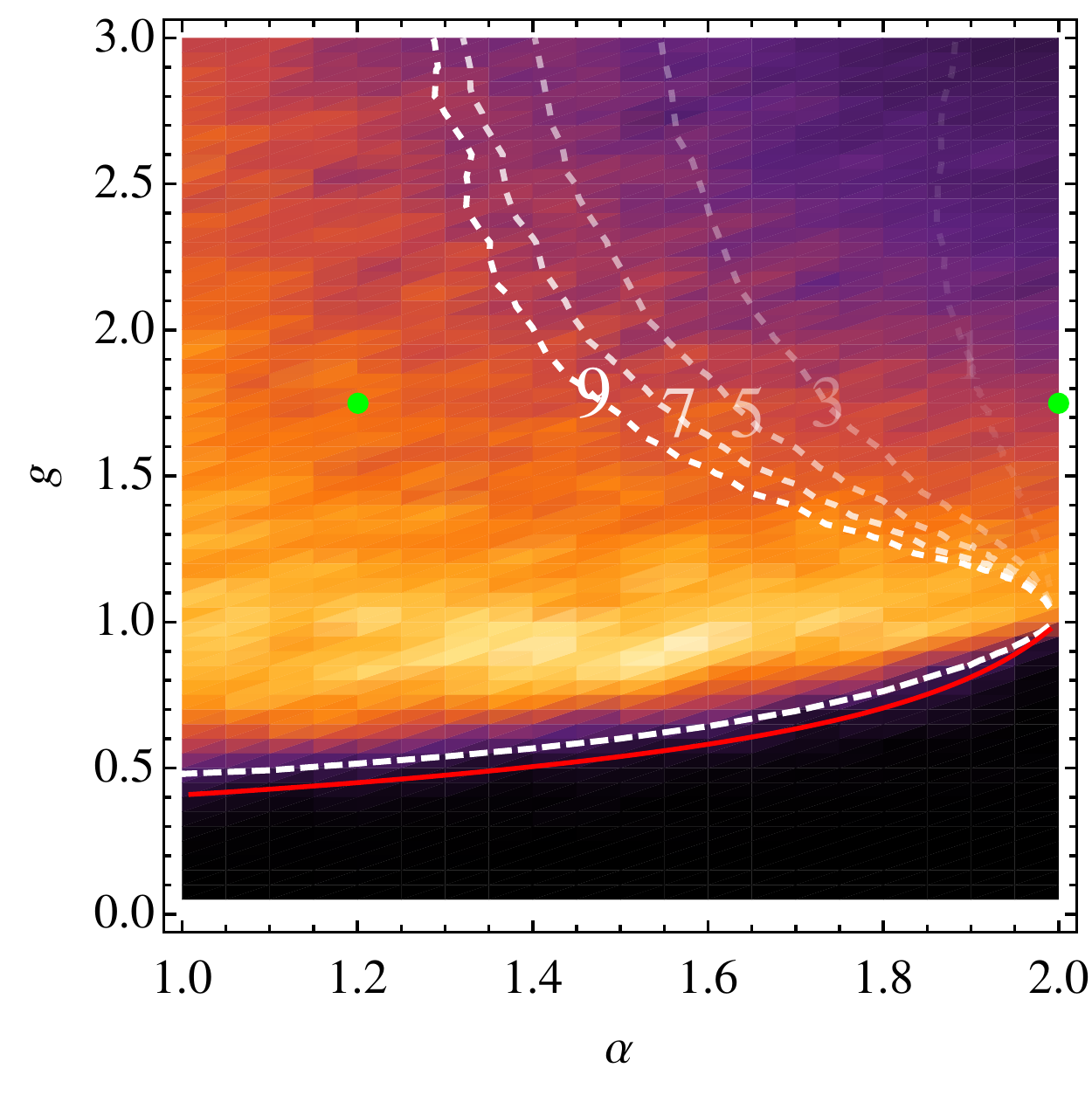}
        \put (0,103) {(c)}
        \end{overpic}
        &
        \begin{tabular}[b]{c}
            \begin{overpic}[width=.5\columnwidth]{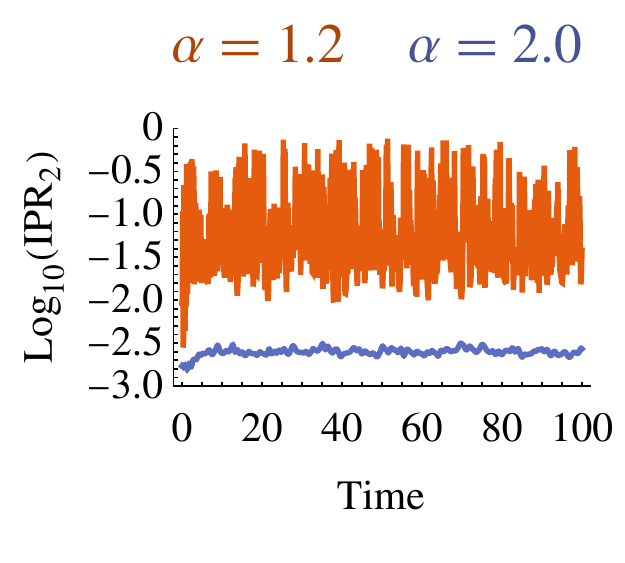}
            \put (0,80) {(d)}
            \end{overpic}
            \\
            \begin{overpic}[width=.5\columnwidth]{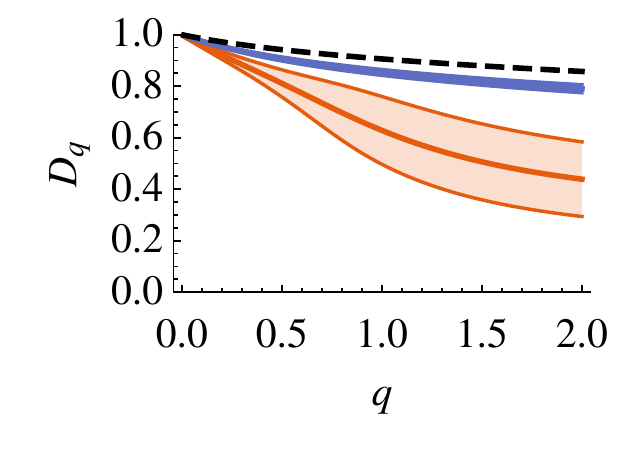}
            \put (0,80) {(e)}
            \end{overpic}
        \end{tabular}
    \end{tabular}
    \caption{The extended Anderson critical regime in heavy-tailed random neural networks.
    (a) Jacobian spectra $J\diag_j(\tanh'(h_j))$ in the stationary state (lines) with characteristic spectral radius $r_{0.01}$ (dashed circle) and their numerical validation (dots).
    (b) Distribution of the half width at quarter-maximum of the local-field autocorrelation for $g=0.75,1.5,2.25,3$ (red, blue, yellow, purple).
    Log-log plot (inset) with comparison power-law line of exponent $-2.0$.
    (c) Phase diagram over heavy-tailed index $\alpha$ and gain $g$.
    The extended Anderson critical regime emerging between the lower and upper dashed lines predicted by Eq.\ (\ref{eq:annealed_average}) (labeled by $n$) corresponds with high relaxation time variance (colored).
    (d) The logarithm of the IPR of the heavy-tailed network activity fluctuation vector (red) varies widely between the asymptotic large-$N$ delocalized ($-\log_{10} N \approx -3$) and localized ($0$) values, while the classical network's activity fluctuations remain delocalized (blue).
    (e) The multifractal dimension $D_q$ estimated via the asymptotic relation $D_q\sim(\log_N \mathrm{IPR}_q)/(1-q)$, shaded to one standard deviation over time.
    $D_q$ estimate for a random delocalized vector on the $N$-sphere (dashed) for comparison.
    }
    \label{fig:network}
\end{figure*}

Figure \ref{fig:network}(a) shows the eigenvalue density $\rho(z)$ of the Jacobian and its numerical validation as a function of eigenvalue modulus $|z|$.
A key characteristic of heavy-tailed neural networks is the infinite spectral radius $r_0$, given by the point at which $y_*=0$, so that the zero fixed point only occurs for zero gain.
However, since the eigenvalue density of the heavy-tailed Jacobian is exponentially suppressed at large radius \cite{Note1,Bordenave2011}, a characteristic spectral radius $r_p$ can be defined by the eigenvalue modulus $|z|$ at which the parameter $y_*$ drops to a fraction $p$ of its value at $z=0$.
Meanwhile, the spectral radius $r_0$ of the Jacobian for classical networks ($\alpha=2$) is finite and is obtained by a spectral analysis of finite-variance block-structured random matrices
\cite{PhysRevLett.97.188104,PhysRevE.85.066116,PhysRevLett.114.088101}.
In this sense, $y_*$ behaves as an order parameter indicating the transition between microscopic and macroscopic numbers of eigenstates relative to system size $N$ at a given eigenvalue modulus $|z|$.
Thus, heavy-tailed networks exhibit a quasi-ordered regime with a fixed point near zero whose magnitude is suppressed due to the microscopic number of eigenstates above the Jacobian stability line $\Re z=1$.
A continuous transition parameterized by $p\ll1$ is then defined by the value of the gain parameter $g$ at which the characteristic spectral radius $r_p=1$, distinguishing a quiescent phase from a chaotic regime (Fig.\ \ref{fig:network}(c), lower dashed line).
This transition is consistent with the point at which neural activity predicted by mean-field theory (Eq.\ (\ref{eq:MFT})) deviates significantly from zero (Fig.\ \ref{fig:network}(c), red line).

\paragraph{An extended critical phase with correlated multifractal modes.---}


Using our cavity approach, we find that all of the right eigenvectors of the heavy-tailed network Jacobian around the stationary state are multifractal for sigmoidal $\phi$, a hallmark of Anderson transitions (see Ref.\ \cite{Note2} for derivation).
The activity of heavy-tailed neural networks is thus dominated by multifractal chaotic fluctuations in contrast to the spatially delocalized chaos appearing in classical models.
This theoretical prediction on heavy-tailed network dynamics is confirmed by simulations of temporal fluctuations of homogeneous and heavy-tailed neural activity (Fig.\ \ref{fig:network}(d--e), red) which are both chaotic at $g=1.75$.
This result is remarkable from a physical standpoint as the activity itself is delocalized due to the bounded activation function $\phi=\tanh$, and does not visibly differ significantly from classical networks with Gaussian dynamics (see Ref.\ \cite{Note2}).

To investigate the behavior of the system's multifractal modes over long timescales, we quantify the extent to which Jacobian eigenvalues $\lambda_i$ corresponding to a given eigenvector change their modulus relative to unity when new samples are chosen from the stationary distribution of neural activity.
We thus consider the Jacobian average
\begin{align}\label{eq:annealed_average}
    \langle(|\lambda_i| - 1)^n\rangle_i
     &= \int_{\mathbb{C}} (|z|-1)^n\rho(z)\,dz 
\end{align}
which penalises small and rewards large eigenvalues to a degree determined by $n$, and we find a region for the gain $g$ adjacent to and above the ordered transition line which is characterised by a greater proportion of eigenvalues away from zero compared to the ordered transition (Fig.\ \ref{fig:network}(c), upper dashed lines).
This allows us to distinguish between an active chaotic region in which unstable fluctuations tend to be quickly suppressed in favour of new fluctuations, and a region of temporally correlated chaotic fluctuations.
The continuous transition between these active and correlated chaotic regimes is parameterized by the annealing strength $n$ in Eq.\ (\ref{eq:annealed_average}).
The theoretically predicted correlated region closes into the well-known critical point at the ordered-chaotic phase transition ($g=1$) for classical rate-based networks ($\alpha=2$), supporting the notion that the extended region of correlated multifractal modes is a critical regime.
This extended critical phase is characterized by a significantly nonzero stationary state and a macroscopic proportion of unstable eigenstates relative to system dimensionality $N$,
which is fundamentally different from
the classical edge of chaos occurring around the zero fixed point when a microscopic proportion of eigenstates crosses the stability line.
Consequently, this extended critical phase remains chaotic rather than existing solely at the edge of a chaotic phase as in classical networks \cite{PhysRevLett.110.118101}.

In summary, the extended critical phase of temporally correlated, spatially multifractal fluctuations provides a demonstration of how various aspects of realistic neural dynamics may be exhibited simultaneously, such as long-range correlations \cite{HE2014480} and low spatial dimensionality relative to system size \cite{MASTROGIUSEPPE2018609}.
This latter property arises from the localization of spatially multifractal fluctuations onto a small number of sites relative to system size (Fig.\ \ref{fig:schematic}).
Such behavior, along with the non-self-averaging properties characteristic of Anderson criticality \cite{PhysRevResearch.3.L032030} which we derived for local-field autocorrelation,
suggests that the timescales across neurons in the heavy-tailed neural network are diverse.
To validate this theoretical prediction on the extended critical region, we compute the relaxation timescales of neural autocorrelations over random networks across heavy-tailed index $\alpha$ and network gain $g$ (Fig.\ \ref{fig:network}(c)).
We find that the extended critical phase is characterized by diverse timescales (Fig.\ \ref{fig:network}(b)): the relaxation timescale distribution is power-law with index $-2.0$, consistent with that seen in cortical memory traces \cite{Bernacchia2011}.
This behavior only exists in the critical regime
\cite{Note2}.

\begin{figure}
    \centering
    \includegraphics[width=\linewidth,trim={0 1cm 0 0},clip]{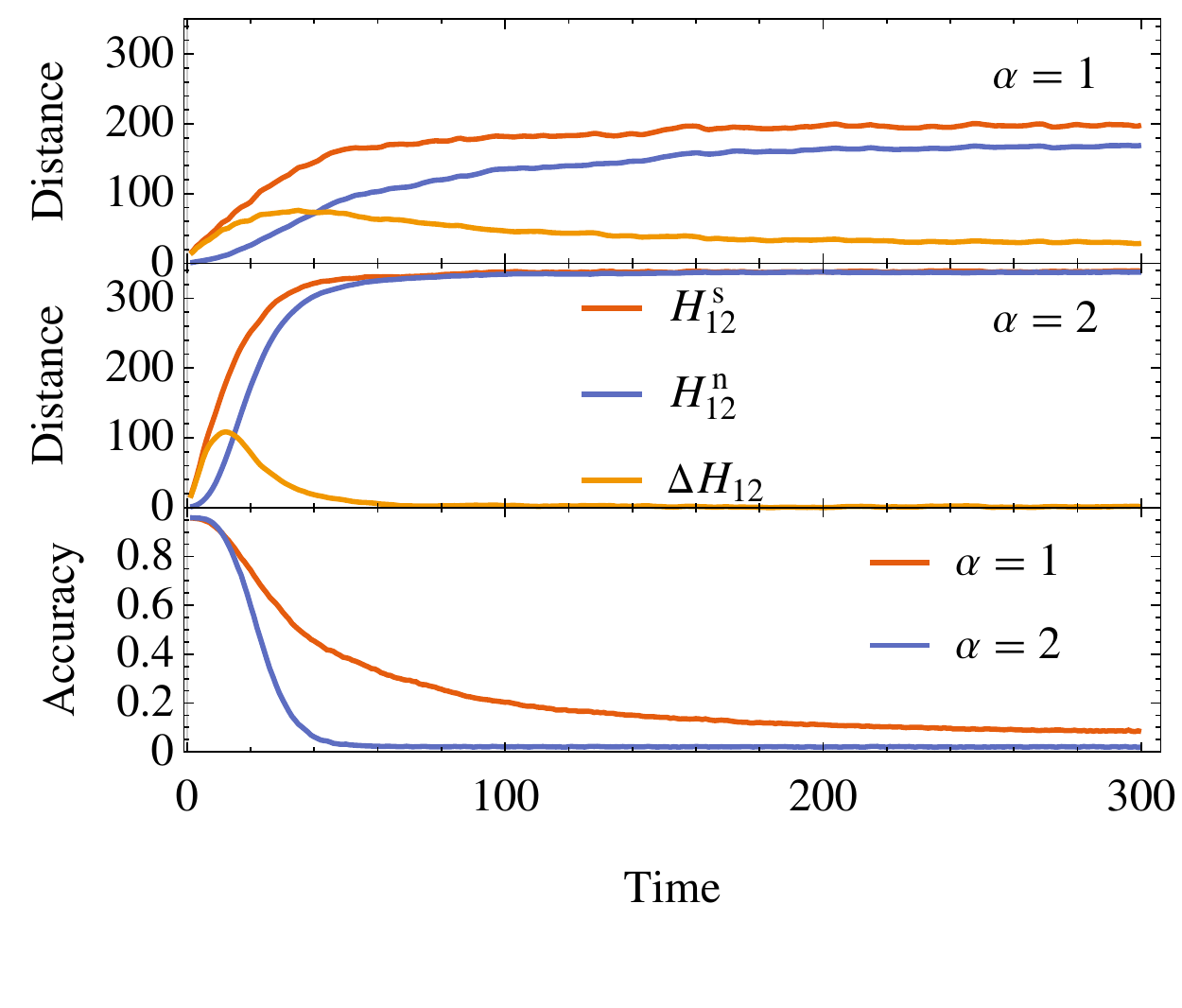}
    \caption{
    Persistent dimensionality expansion in heavy-tailed random neural networks.
    Top, middle panels: The evolution of average signal and noise distances in a rate-based model with $N=250$ and $g=3$, as measured by the Hamming distance $H_{12}:=\|\phi(h_i^{(1)})-\phi(h_i^{(2)})\|_i^2$ between states $\phi(h_i^{(1)})$ and $\phi(h_i^{(2)})$.
    Bottom panel: Classification accuracy of heavy-tailed (red) and classical (blue) rate-based neural networks.}
    \label{fig:classifier_accuracy}
\end{figure}

\paragraph{Persistent reservoir computing.---}

To explore the computational implications of the extended critical regime of correlated multifractal chaos, we consider a reservoir computing task described in \cite{PhysRevX.11.021064} exploiting the chaotic dimensionality expansion of neural representations $\Delta H_{12} = H_{12}^{(s)} - H_{12}^{(n)}$ between signal ($H_{12}^{(s)}$) and noise ($H_{12}^{(n)}$) distances (see Ref.\ \cite{Note2} for setup details).
Heavy-tailed networks make use of the temporal correlations of multifractal Jacobian eigenvectors above the stability line to enact a persistent form of real-time computation in a reservoir computing context.
Because the extended critical regime is chaotic in heavy-tailed networks with a significantly nonzero fixed point, this regime is able to perform dimensionality expansion on its input in contrast to the classical edge of chaos which resides around the zero fixed point.
At the same time, the correlated multifractal Jacobian eigenvectors work to hold off the onset of mixing (Fig.\ \ref{fig:classifier_accuracy}, yellow) to perform persistent chaotic dimensionality expansion on its input, allowing the computed result to remain in the system and thus the classification performance to stay above the baseline ($0.02$) for a longer period of time (Fig.\ \ref{fig:classifier_accuracy}, bottom).
The extended critical regime of correlated multifractal chaos is thus able to produce efficient neural representations balancing the dimensional compression of stimuli (Fig.\ \ref{fig:classifier_accuracy} top, red and blue), which is useful for generalization \cite{Stringer2019}, and the separation of stimuli, in order to enact a form of persistent real-time computation.

\paragraph{Discussion.---}


Our theory rigorously demonstrates that heterogeneous, heavy-tailed connectivity can endow neural circuits
with Anderson criticality over an extended parameter region, thus eliminating the fine-tuning needed in homogeneously connected neural networks.
The Anderson criticality is characterized by correlated, low-dimensional fluctuations
\cite{HE2014480,MASTROGIUSEPPE2018609}
and a diverse reservoir of timescales \cite{Bernacchia2011} as observed in biological neural systems; these observations otherwise remain unexplained in a unified manner in conventional theories.
Moreover, the extended Anderson criticality provides a unique mechanism for combining robust real-time computation with long-term memory of the computed output.
Both homogeneous and heavy-tailed networks utilize chaos to enhance the separation of inputs for linear classification, but after a transient period of high separability, mixing dominates and erases the computed output from the classical homogeneous system.
The recent observations on the ubiquity of heavy-tailed coupling in pretrained deep neural networks \cite{Martin2021} suggest that our theory would be powerful for revealing the shared dynamical principles for persistent computation in both biological and artificial neural networks.

By making a novel link between Anderson criticality and the highly fluctuating complex dynamics of neural networks, our results suggest that complex systems operating over multiple scales should display a degree of multifractality at some level of fluctuations of system activity, even when the activity itself is bounded due to physical constraints.
Multifractal phenomena have indeed been seen in a wide variety of natural systems such as turbulence \cite{Benzi_1984}.
In statistical and condensed matter physics, the Anderson transition appears with multifractality around the boundary of two phases characterised by localized and delocalized eigenvectors \cite{RevModPhys.80.1355,PhysRevLett.116.010601}.
Our heavy-tailed neural network may thus be viewed as an inversion of these canonical models: a structurally extended Anderson regime $0<\alpha<2$ is bounded by delocalized Jacobian modes appearing in classical homogeneous networks with $\alpha=2$, and localized eigenstates appearing in the limit $\alpha\to0$ corresponding to sparse matrices and directed random graphs with a small average degree \cite{PhysRevLett.126.040604}.
Hence our random matrix theory could be applied to understand how complex dynamics emerge in physical systems.

\bibliography{main}




\section*{Supplementary material}

# Supplemental Material: Extended Anderson Criticality in Heavy-Tailed Neural Networks

Asem Wardak and Pulin Gong*
*School of Physics, University of Sydney, New South Wales 2006, Australia*
(Dated: February 11, 2022)

In this Supplemental Material, we expand on the mathematical results underpinning this Letter. We begin with a review of non-Hermitian random matrix theory and the bipartisation method. We then present new theoretical results by developing a cavity approach on column-structured non-Hermitian heavy-tailed random matrices, allowing us to analytically treat the Jacobian matrices in the main text. Finally, we present some dynamical Lévy mean-field derivations which apply to heavy-tailed random neural networks, along with additional background information on some of the results of the Letter.

In the following, the notation $\langle f(x_i) \rangle_i = (1/N) \sum_{i=1}^{N} f(x_i)$ applies in the limit of large matrix size.

## CONTENTS

## I. ELEMENTS OF NON-HERMITIAN RANDOM MATRIX THEORY

In this section we review the random matrix theory required for understanding the cavity method [1, 2]. The eigenvalues $\lambda_k$ of an $N \times N$ matrix $A$ are described by the spectral density $\rho_A(z) := \langle \delta(z - \lambda_k) \rangle_k$, which can be computed using the resolvent $G_A(z) := (A - zI)^{-1}$ via the relation $\pi \rho_A(z) = -\partial_{\bar{z}} \langle G_A(z)_{ii} \rangle_i$, where $\partial_z := (\partial_x - i\partial_y)/2$ for $z = x + iy$. Writing the resolvent in its canonical form

$$G_A(z) = \sum_{k=1}^{N} \frac{|v_k\rangle\langle u_k|}{\lambda_k - z} \qquad (1)$$

where $A|v_k\rangle = \lambda_k |v_k\rangle$ and $\langle u_k| A = \lambda_k \langle u_k|$ are the right and left eigenvectors of $A$, one may evaluate the resolvent in the neighbourhood of an eigenvalue to deduce the correlations between the corresponding left and right eigenvectors.

---

* pulin.gong@sydney.edu.au

## A. Hermitian matrices

This canonical form simplifies for a Hermitian matrix $H$ to

$$G_H(\lambda + i\epsilon)_{ii} = \sum_{k=1}^{N} \frac{|\langle i|v_k\rangle|^2}{\lambda_k - \lambda - i\epsilon} = \sum_{k=1}^{N} \frac{|\langle i|v_k\rangle|^2(\lambda_k - \lambda + i\epsilon)}{(\lambda_k - \lambda)^2 + \epsilon^2} \xrightarrow{\epsilon \to 0} G_H(\lambda)_{ii} + i\pi N\rho(\lambda)|\langle i|v(\lambda)\rangle|^2 , \quad (2)$$

yielding a simpler form of the spectral density $\pi\rho_H(\lambda) = \lim_{\epsilon \to 0}\langle \operatorname{Im} G_H(\lambda + i\epsilon)_{ii}\rangle_i$ along with the inverse participation ratio $\operatorname{IPR}_q := \sum_i |\langle i|v\rangle|^{2q}$ used to characterise the localisation of the eigenvectors,

$$\operatorname{IPR}_{q,H}(\lambda) = N^{1-q} \lim_{\epsilon \to 0} \frac{\langle (\operatorname{Im} G_H(\lambda + i\epsilon)_{ii})^q \rangle_i}{\langle \operatorname{Im} G_H(\lambda + i\epsilon)_{ii}\rangle_i^q} . \quad (3)$$

Denoting the fractal dimension $D_q$ by $\operatorname{IPR}_q \sim N^{D_q(1-q)}$ [3], delocalised and localised eigenvectors are characterised by $D_q$ equalling unity and zero respectively, while multifractal eigenvectors are characterised by $D_q$ being a nontrivial function of $q$. To determine the exact value of the IPR when the eigenvectors are localised, another technique may be used by taking the lowest-order expansion of the $q$-th power of the sum when $q > 1$,

$$|G_H(\lambda + i\epsilon)_{ii}|^q = \sum_k \frac{|\langle i|v_k\rangle|^{2q}}{|\lambda_k - \lambda - i\epsilon|^q} + \cdots = \sum_k \frac{|\langle i|v_k\rangle|^{2q}}{((\lambda_k - \lambda)^2 + \epsilon^2)^{q/2}} + \cdots = \sum_k |\langle i|v_k\rangle|^{2q} c_{q,\epsilon} \delta(\lambda_k - \lambda) \quad (4)$$

$$= N\rho(\lambda) c_{q,\epsilon} |\langle j|v(\lambda)\rangle|^{2q} , \quad (5)$$

where $c_{q,\epsilon} := \int_{-\infty}^{\infty} dx/(x^2 + \epsilon^2)^{q/2} = \sqrt{\pi}\epsilon^{1-q}\Gamma(\frac{q-1}{2})/\Gamma(\frac{q}{2})$, yielding for $q > 1$ (cf. Eq. (5) of [4])

$$\operatorname{IPR}_{q,H}(\lambda) = \lim_{\epsilon \to 0} \frac{1}{\rho(\lambda) c_{q,\epsilon}} \langle |G_H(\lambda + i\epsilon)_{ii}|^q \rangle_i . \quad (6)$$

## B. Bipartised matrix: quaternionic resolvent

However, for non-Hermitian matrices whose bulk spectra have a nonempty interior, the resolvent cannot be evaluated in any neighbourhood of these eigenvalues due to the presence of nearby poles. Moreover, one cannot separate the left and right eigenvector elements in the canonical form of the resolvent. These issues can be alleviated by using the resolvent of an auxiliary matrix of twice the dimensionality via a bipartisation [1, 5] or Hermitisation [2, 6, 7] method. To see this, observe that the $2N \times 2N$ Hermitian matrix

$$H(z, \eta) = \begin{pmatrix} -\eta I_N & A - zI_N \\ A^\dagger - \bar{z}I_N & -\eta I_N \end{pmatrix} \quad (7)$$

has inverse

$$H^{-1}(z, \eta) = G_{H(z,0)}(\eta) = \begin{pmatrix} \eta X & X(A - zI_N) \\ (A^\dagger - \bar{z}I_N)X & \eta Y \end{pmatrix} \quad (8)$$

where $X = ((A-z)(A^\dagger - \bar{z}) - \eta^2)^{-1}$ and $Y = ((A^\dagger - \bar{z})(A - z) - \eta^2)^{-1}$. The lower left block of $H^{-1}(z, \eta)$ is then equal to $G_A(z) + O(|\eta|^2)$ for small $|\eta|$, so that the spectral density of $A$ satisfies $\pi\rho_A(z) = -\lim_{|\eta| \to 0} \partial_{\bar{z}}\langle H^{-1}(z, \eta)_{(i+N),i}\rangle_i$. Meanwhile, for all $c_1, c_2 \in \mathbb{C}$,

$$H(\lambda_k(A), \eta)\begin{pmatrix} c_1|u_k(A)\rangle \\ c_2|v_k(A)\rangle \end{pmatrix} = -\eta \begin{pmatrix} c_1|u_k(A)\rangle \\ c_2|v_k(A)\rangle \end{pmatrix} . \quad (9)$$

Thus the left and right eigenvectors of $A$ are given by the diagonal elements of the top and bottom diagonal blocks of $H^{-1}(z, \eta)$ respectively, while the eigenvalues of $A$ are given by the diagonal of either off-diagonal block of $H^{-1}(z, \eta)$. Permuting the nodes of $H$ yields an $N \times N$ matrix $\mathbf{A}$ with quaternionic elements

$$\mathbf{A}_{ij} = \begin{pmatrix} 0 & A_{ij} \\ A_{ji} & 0 \end{pmatrix} , \qquad U(z, \eta) = \begin{pmatrix} \eta & z \\ \bar{z} & \eta \end{pmatrix} . \quad (10)$$

The eigenvalues and eigenvectors of $A$ can then be recovered with a quaternionic notion of the resolvent, $\mathbf{G}_\mathbf{A}(U) := (\mathbf{A} - U \otimes I_N)^{-1}$, which is related to the usual resolvent via $\mathbf{G}_\mathbf{A}(U(z,\eta)) = G_{\mathbf{A}-U(z,0)\otimes I_N}(\eta)$. To see this, denote the diagonal elements of the quaternionic resolvent by

$$\mathbf{G}_\mathbf{A}(U)_{ii} =: \begin{pmatrix} a_i(z,\eta) & b_i(z,\eta) \\ b'_i(z,\eta) & c_i(z,\eta) \end{pmatrix} . \qquad (11)$$

If $\eta \in i\mathbb{R}_+$, then $a_i, c_i \in i\mathbb{R}_+$ (i.e. $a_i = i|a_i|$), and $b'_i = \overline{b_i}$ (Lemma 2.2 of [5]). Then $\pi \rho_A(z) = -\lim_{t\to 0}\langle \partial_z b_i(z,it)\rangle_i$ and the inverse participation ratios of the left and right eigenvectors of $A$ are

$$\mathrm{IPR}^{(l)}_{q,A}(z) = N^{1-q} \lim_{t\to 0} \frac{\langle (\mathrm{Im}\, a_i(z,it))^q\rangle_i}{\langle \mathrm{Im}\, a_i(z,it)\rangle_i^q} , \qquad \mathrm{IPR}^{(r)}_{q,A}(z) = N^{1-q} \lim_{t\to 0} \frac{\langle (\mathrm{Im}\, c_i(z,it))^q\rangle_i}{\langle \mathrm{Im}\, c_i(z,it)\rangle_i^q} . \qquad (12)$$

## II. THE CAVITY METHOD FOR COLUMN-STRUCTURED NON-HERMITIAN HEAVY-TAILED RANDOM MATRICES

In this section we introduce a new cavity method for obtaining the resolvent of random matrices obtained by scaling the columns of an asymmetric heavy-tailed random matrix by a set of values, deterministic or random, such as those appearing in Jacobian matrices of heavy-tailed random neural networks. This will be used to compute the spectral density and eigenvector localisation of the Jacobian matrices of heavy-tailed neural networks considered in the main text.

To compute the diagonal of the resolvent for matrices with scalar or vector entries, one may employ the blockwise inversion formula

$$\begin{pmatrix} a & b \\ c & d \end{pmatrix}^{-1} = \begin{pmatrix} s_d^{-1} & -s_d^{-1} b d^{-1} \\ -d^{-1} c s_d^{-1} & s_a^{-1} \end{pmatrix} , \qquad (13)$$

where $s_d = a - bd^{-1}c$ is the Schur complement of $d$ and $s_a = d - ca^{-1}b$ is the Schur complement of $a$. The diagonal elements of the quaternionic resolvent of $\mathbf{A}$ thus satisfy the recursion relation

$$\mathbf{G}_\mathbf{A}(U)_{ii} = \left( \mathbf{A}_{ii} - U - \sum_{j,k\neq i} \mathbf{A}_{ij} \mathbf{G}^{(i)}_{jk} \mathbf{A}_{ki} \right)^{-1} , \qquad (14)$$

where $\mathbf{G}^{(i)}_{jk} = \mathbf{G}_{\mathbf{A}^{(i)}}(U)_{jk}$ is the resolvent of the adjacency matrix $\mathbf{A}^{(i)}$ of the graph with node $i$ removed.

### A. Cavity method for scalar and vector treelike matrices

If $A$ is the adjacency matrix of a weighted tree, then so too is $\mathbf{A}$ and the sum is restricted to $j,k \in \partial_i$, where $\partial_i$ are the neighbouring nodes of $i$. Moreover, since $\mathbf{A}^{(i)}$ is the adjacency matrix of a forest of isolated trees with roots $j \in \partial_i$, the relation $j \neq k$ implies that $j,k$ belong to distinct trees in $\mathbf{A}^{(i)}$ and $\mathbf{G}^{(i)}_{jk} = 0$ (seen by permuting the blockwise inversion formula with $b,c = 0$). Thus

$$\mathbf{G}_\mathbf{A}(U)_{ii} = \left( \mathbf{A}_{ii} - U - \sum_{j\in\partial_i} \mathbf{A}_{ij} \mathbf{G}^{(i)}_{jj} \mathbf{A}_{ji} \right)^{-1} . \qquad (15)$$

However, $G_A(z)_{jj} \neq G_{A^{(i)}}(z)_{jj}$ in general. Removing another node from $A^{(i)}$ brings the recursion to a second step:

$$\mathbf{G}^{(i)}_{jj} = \left( \mathbf{A}_{jj} - U - \sum_{k\in\partial_j\setminus i} \mathbf{A}_{jk} \mathbf{G}^{(i,j)}_{kk} \mathbf{A}_{kj} \right)^{-1} , \qquad (16)$$

where $j \in \partial_i$. To close the recursion observe that $\mathbf{G}^{(i,j)}_{kk} = \mathbf{G}^{(j)}_{kk}$ in Fig. 1, and the subtrees rooted at $j$ and $k$ are

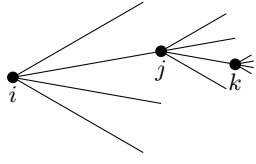

FIG. 1. The cavity method.

statistically identical over $\partial_i$ and $\partial_j \setminus i$ respectively. Thus $(\mathbf{G}_{jj}^{(i)})_{j\in\partial_i}$ and $(\mathbf{G}_{kk}^{(j')})_{k\in\partial_{j'}\setminus i}$ have the same distribution for all $j' \in \partial_i$. In particular,

$$\langle f(\mathbf{G}_{jj}^{(i)})\rangle_{j\in\partial_i} = \langle f(\mathbf{G}_{kk}^{(j)})\rangle_{k\in\partial_j\setminus i} \tag{17}$$

for any function $f$.

The diagonal elements $\mathbf{G}_\mathbf{A}(U)_{ii}$ of the quaternionic resolvent are thus obtained by solving the recursive distributional equation

$$\begin{pmatrix} a_j^{(i)} & b_j^{(i)} \\ \bar{b}_j^{(i)} & c_j^{(i)} \end{pmatrix} = -\left(\begin{pmatrix} \eta & z - A_{jj} \\ \bar{z} - \overline{A_{jj}} & \eta \end{pmatrix} + \sum_{k\in\partial_j\setminus i} \begin{pmatrix} |A_{jk}|^2 c_k^{(j)} & A_{jk}A_{kj}\bar{b}_k^{(j)} \\ \overline{A_{kj}A_{jk}}b_k^{(j)} & |A_{kj}|^2 a_k^{(j)} \end{pmatrix}\right)^{-1}. \tag{18}$$

### B. Spectral theory of column-structured non-Hermitian heavy-tailed random matrices

Consider the case $\eta = it \in i\mathbb{R}_+$ and let $A_{ij} = J_{ij}\chi_j$ be the column-structured non-Hermitian heavy-tailed random matrix defined in the main text. Assume that $\langle|\chi_i|^\alpha\rangle_i$ is finite and, if $\chi$ is a random variable, that $\chi_j$ is independent of $J_{ij}$. Since $J$ converges to a random operator on a tree for large $N$ [5], and

$$p_{J_{ij}^2}(x) \overset{x\to\infty^+}{\sim} \frac{c_\alpha}{2n|x|^{1+\alpha/2}} = \frac{c_\alpha}{2c_{\alpha/2}} \frac{c_{\alpha/2}}{n|x|^{1+\alpha/2}} \tag{19}$$

for $x > 0$, the generalised central limit theorem gives the recursive distributional equation

$$\begin{pmatrix} a_j^{(i)} & b_j^{(i)} \\ \bar{b}_j^{(i)} & c_j^{(i)} \end{pmatrix} = \frac{1}{D}\begin{pmatrix} i|\chi_j|^2\langle|a_k^{(j)}|^{\alpha/2}\rangle_{k\in\partial_j\setminus i}^{2/\alpha} S_a^{(j)} & -z \\ -\bar{z} & i\langle|\chi_k|^\alpha|c_k^{(j)}|^{\alpha/2}\rangle_{k\in\partial_j\setminus i}^{2/\alpha} S_c^{(j)} \end{pmatrix} \tag{20}$$

in the limit $t \to 0$, where $S_a^{(j)}, S_c^{(j)} \sim L(\alpha/2, 1, 0, C_\alpha/4C_{\alpha/2})$ are independent (as in the operator limit $A_{ji} = 0$ whenever $A_{ij}$ is not [5]), and

$$D = |z|^2 + |\chi_j|^2\langle|a_k^{(j)}|^{\alpha/2}\rangle_{k\in\partial_j\setminus i}^{2/\alpha} S_a^{(j)} \langle|\chi_k|^\alpha|c_k^{(j)}|^{\alpha/2}\rangle_{k\in\partial_j\setminus i}^{2/\alpha} S_c^{(j)}. \tag{21}$$

The equivalence of the distributions $(\mathbf{G}_{jj}^{(i)})_{j\in\partial_i}$ and $(\mathbf{G}_{kk}^{(j')})_{k\in\partial_{j'}\setminus i}$ for any node $j' \in \partial_i$ gives the following coupled equations for $\langle|a_k^{(j)}|^{\alpha/2}\rangle_{k\in\partial_j\setminus i}$ and $\langle|\chi_k|^\alpha|c_k^{(j)}|^{\alpha/2}\rangle_{k\in\partial_j\setminus i}$,

$$\langle|a_k^{(j)}|^{\alpha/2}\rangle_{k\in\partial_j\setminus i} = \langle|a_j^{(i)}|^{\alpha/2}\rangle_{j\in\partial_i} = \left\langle\left(\frac{|\chi_j|^2\langle|a_k^{(j)}|^{\alpha/2}\rangle_{k\in\partial_j\setminus i}^{2/\alpha} S_a^{(j)}}{|z|^2 + |\chi_j|^2\langle|a_k^{(j)}|^{\alpha/2}\rangle_{k\in\partial_j\setminus i}^{2/\alpha} S_a^{(j)} \langle|\chi_k|^\alpha|c_k^{(j)}|^{\alpha/2}\rangle_{k\in\partial_j\setminus i}^{2/\alpha} S_c^{(j)}}\right)^{\alpha/2}\right\rangle_{j\in\partial_i}, \tag{22}$$

$$\langle|\chi_k|^\alpha|c_k^{(j)}|^{\alpha/2}\rangle_{k\in\partial_j\setminus i} = \langle|\chi_j|^\alpha|c_j^{(i)}|^{\alpha/2}\rangle_{j\in\partial_i} = \left\langle\left(\frac{|\chi_j|^2\langle|\chi_k|^\alpha|c_k^{(j)}|^{\alpha/2}\rangle_{k\in\partial_j\setminus i}^{2/\alpha} S_c^{(j)}}{|z|^2 + |\chi_j|^2\langle|a_k^{(j)}|^{\alpha/2}\rangle_{k\in\partial_j\setminus i}^{2/\alpha} S_a^{(j)} \langle|\chi_k|^\alpha|c_k^{(j)}|^{\alpha/2}\rangle_{k\in\partial_j\setminus i}^{2/\alpha} S_c^{(j)}}\right)^{\alpha/2}\right\rangle_{j\in\partial_i}. \tag{23}$$

The invariance of these two equations under the symmetry $\langle|a_k^{(j)}|^{\alpha/2}\rangle_{k\in\partial_j\setminus i} \leftrightarrow \langle|\chi_k|^\alpha|c_k^{(j)}|^{\alpha/2}\rangle_{k\in\partial_j\setminus i}$ yields a solution on $(0,\infty)$ given by $\langle|a_k^{(j)}|^{\alpha/2}\rangle_{k\in\partial_j\setminus i} = \langle|\chi_k|^\alpha|c_k^{(j)}|^{\alpha/2}\rangle_{k\in\partial_j\setminus i} =: y_*$, where[1]

$$y_*^{\alpha/2} = \left\langle \left(\frac{y_*|\chi_j|^2 S}{|z|^2 + y_*^2|\chi_j|^2 SS'}\right)^{\alpha/2} \right\rangle_{j\in\partial_i} . \tag{24}$$

Hence the diagonal elements of the quaternionic resolvent are

$$\begin{pmatrix} a_i & b_i \\ \bar{b}_i & c_i \end{pmatrix} = \lim_{t\to 0} \frac{1}{|z|^2 + (t+y_*|\chi_i|^2 S)(t+y_* S')} \begin{pmatrix} it + iy_*|\chi_i|^2 S & -z \\ -\bar{z} & it + iy_* S' \end{pmatrix} . \tag{25}$$

This gives

$$\rho(z) = \frac{1}{\pi}(y_*^2 - 2|z|^2 y_* \partial_{|z|^2} y_*) \left\langle \frac{|\chi_i|^2 SS'}{(|z|^2 + |\chi_i|^2 y_*^2 SS')^2} \right\rangle_i \tag{26}$$

for the eigenvalues of $A$ and

$$\text{IPR}_q^{(l)}(z) = c_q N^{1-q} \frac{\left\langle \left(\frac{y_*|\chi_i|^2 S}{|z|^2 + y_*^2|\chi_i|^2 SS'}\right)^q \right\rangle_i}{\left\langle \frac{y_*|\chi_i|^2 S}{|z|^2 + y_*^2|\chi_i|^2 SS'} \right\rangle_i^q} , \qquad \text{IPR}_q^{(r)}(z) = c_q N^{1-q} \frac{\left\langle \left(\frac{y_* S}{|z|^2 + y_*^2|\chi_i|^2 SS'}\right)^q \right\rangle_i}{\left\langle \frac{y_* S}{|z|^2 + y_*^2|\chi_i|^2 SS'} \right\rangle_i^q} , \tag{27}$$

for the left and right IPRs, where $c_q$ is a constant factor determined by the local eigenvalue statistics. To compute $\partial_{|z|^2} y_*$, use the implicit function theorem. Defining

$$G(y,r) = \left|\frac{|\chi_i|^2 S}{r + y^2 |\chi_i|^2 SS'}\right|^{\alpha/2} \tag{28}$$

yields $\partial_{|z|^2} y_* = -\langle \partial_r G(y_*, |z|^2)\rangle / \langle \partial_y G(y_*, |z|^2)\rangle$, where

$$\partial_r G(y,r) = \frac{-(\alpha/2)|\chi_i|^\alpha S^{\alpha/2}}{(r + y^2|\chi_i|^2 SS')^{\alpha/2+1}} , \qquad \partial_y G(y,r) = 2y|\chi_i|^2 SS' \partial_r G(y,r) . \tag{29}$$

The density of eigenvalues at $z = 0$ is

$$\rho(0) = y_*^{-2}|_{z=0} \langle |\chi_i|^{-2}\rangle_i \langle S^{-1}\rangle^2 , \tag{30}$$

which may be infinite due to $\chi$. To obtain $y_*^{-2}|_{z=0}$, observe that $y|_{z=0}^\alpha = \langle S^{-\alpha/2}\rangle$ along with the fact that if $S_D \sim L(\beta, 1, 0, D)$ with $\beta < 1$, then

$$\langle S_D^{-\beta}\rangle = \frac{\cos(\beta\pi/2)}{D\Gamma(1+\beta)} . \tag{31}$$

Noting the constraint $\langle|\chi_i|^\alpha\rangle_i < \infty$, it can be shown using similar techniques to [5] that the spectral radius $r_0$ is in general infinite, and that the eigenvalue density is exponentially suppressed at large modulus in stark contrast to the power-law distribution of singular values.

An analogous derivation for the classical finite-variance case $\alpha = 2$ yields the above results with the transformation $S \mapsto 1$. In this case, the eigenvalue density for column-structured matrices of finite-variance elements is recovered along with the spectral radius $\langle|\chi_i|^2\rangle_i^{1/2}$ as known previously [8–10], while a new result on the localisation of column-structured matrices of finite-variance elements arises from the analogue to Eq. (27).

---

[1] In the case of block-structured connectivity $gJ_{ij} \mapsto g_{c_i c_j} J_{ij}$, which describes multipopulation neural networks with varying cell types, this fixed-point equation for $y_*$ is replaced by a set of coupled equations $y_{*,c}$ over the cell types.

## C. Eigenvector localisation for various column structures

Let us now characterise the localisation properties of the IPR for various distributions of $\chi$. In the case of non-Hermitian heavy-tailed random matrices [5] ($\chi = 1$), the left and right eigenvectors are delocalised and can be directly evaluated using Eq. (27). However, since many common sigmoidal transfer functions are bounded or flatten out for large inputs, and local-field stationary distributions have power-law tails, $\chi = g\phi'(h)$ is more likely to have significant mass near zero.

First consider the test case[2] $\mathbb{P}(\chi_i = 0) > 0$ and consider the sum $\sum_i |c_i(z,i0)|^q$ using Eq. (25),

$$\sum_{i=1}^N |c_i(z,i0)|^q = \sum_{i=1}^N (\operatorname{Im} c_i(z,i0))^q = \sum_{i=1}^N \left|\frac{y_* S'}{|z|^2 + y_*^2 |\chi_i|^2 SS'}\right|^q. \tag{32}$$

In this sum, $\mathbb{P}(\chi_i = 0)N$ terms are of the form $(y_* S'/|z|^2)^q$, which has the asymptotic tail of an $\alpha/2q$-stable random variable if $\alpha/2q < 1$, and has finite mean otherwise. Hence the sum scales as $N^{2q/\alpha}$ if $q > \alpha/2$, and as $N$ otherwise. Thus the right IPR scales as $N^{2q/\alpha}/(N^{2/\alpha})^q = N^0$ when $q > \alpha/2$ and $N/(N^{2/\alpha})^q = N^{1-2q/\alpha}$ otherwise. As a result the multifractal dimension $D_q = 0$ when $q > \alpha/2$, and $D_q = (1 - 2q/\alpha)/(1 - q)$ otherwise, i.e. the right eigenvectors are multifractal. Meanwhile, the left eigenvector remains delocalised and can be directly evaluated using Eq. (27).

Now consider the general case where the distribution of $\chi$ has a certain asymptotic scaling around zero. Suppose the quantity $\mathbb{P}(|\chi|^2 S < \epsilon)$ is known for small $\epsilon$. Of the sum $\sum_i |c_i(z,it)|^q$, $\mathbb{P}(y_* |\chi|^2 S < t)N$ terms are in the interval

$$\left(\left|\frac{t + y_* S'}{|z|^2 + (t+t)(t+y_* S')}\right|^q, \left|\frac{t + y_* S'}{|z|^2 + t(t+y_* S')}\right|^q\right). \tag{33}$$

Of these terms, $\mathbb{P}(y_*|\chi|^2 S < t)N\mathbb{P}(ty_* S' > |z|^2) = \mathbb{P}(|\chi|^2 S < t/y_*)N(|z|^2/ty_*)^{-\alpha/2}$ terms have the scaling $1/t^q$. Noting that the stability of the average of the diagonal elements of the scalar-valued resolvent upon the addition of a small imaginary part is classically used as a criterion for localisation for the Anderson transition [11], this average for the right eigenvector elements scales over large $N$ and small $t$ as $\mathbb{P}(|\chi|^2 S < t)t^{\alpha/2-1}N^0$. If this average remains finite,[3]

$$\mathbb{P}(|\chi|^2 S < 1/N) = O(N^{\alpha/2-1}) \qquad \text{(deloc.)} \tag{34}$$

then the state is considered delocalised (for example, if the density of $|\chi|^2$ at zero is finite or if $\mathbb{P}(|\chi|^2 S < t) \sim t^a$ with $a > 1 - \alpha/2 > 0$). However, if this average is unstable for small $t$,

$$\mathbb{P}(|\chi|^2 S < 1/N) \neq O(N^{\alpha/2-1}) \qquad \text{(not deloc.)} \tag{35}$$

(e.g. $\mathbb{P}(|\chi|^2 S < t) \sim t^a$ with $0 < a < 1 - \alpha/2$ or $|\chi|^2$ has sufficient mass near zero) then the state is localised to some degree as $\operatorname{Im} G_{\mathbf{A} - U(z,0) \otimes I_N}(it)_{(2i),(2i)}$ diverges when $t \to 0$. In this case the second IPR definition yields the right IPR scaling to be $t^0 N^0$ for $q > 1$. Since any configuration of $\chi$ yields a greater or equal delocalisation than the test case of a delta function at zero, the right eigenvector is multifractal in this case and the multifractal dimension satisfies

$$D_q \geq \frac{1 - 2q/\alpha}{1 - q} \qquad \text{when } q < \alpha/2, \tag{36}$$

$$D_q \geq 0 \qquad \text{when } \alpha/2 < q < 1, \tag{37}$$

$$D_q = 0 \qquad \text{when } q > 1. \tag{38}$$

For what transfer functions $\phi$ does the multifractality condition $\mathbb{P}(\phi'(h)^2 S < 1/N) \neq O(N^{\alpha/2-1})$ hold? If $\phi$ is sigmoidal (bounded, one inflection point, non-negative derivative), we can observe that asymptotically for large $N$

$$\mathbb{P}(\phi'(h)^2 S < 1/N) > \mathbb{P}(\phi'(h)^2 < 1/N)\mathbb{P}(S < 1) \tag{39}$$

$$\sim \mathbb{P}(\phi'(h) < 1/\sqrt{N}) \tag{40}$$

$$= \mathbb{P}(h > (\phi')^{-1}(1/\sqrt{N})) \tag{41}$$

$$\sim ((\phi')^{-1}(1/\sqrt{N}))^{-\alpha} \tag{42}$$

---

[2] This case arises, for example, when $\phi = \Theta(x - \theta)$ is a binary activation function or has a hard threshold.
[3] Note that $\mathbb{P}(|\chi|^2 S < 1/N)$ is monotonically decreasing as $N \to \infty$, and that the exponent of $N^{\alpha/2-1}$ is negative.

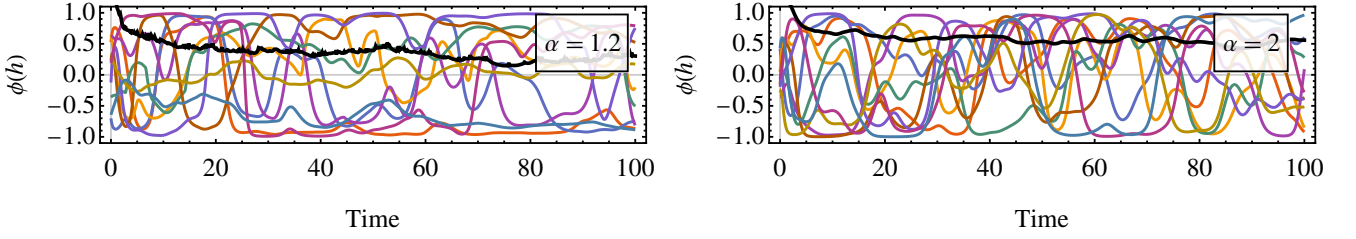

FIG. 2. Comparison of spontaneous activity between homogeneous and heavy-tailed networks for several representative neurons (colours) with $N = 2000$ and $g = 1.75$. Magnitude of the temporal fluctuations of network activity (black) measured by the mean absolute value of the time derivative of the neural activity vector.

The derivatives of bounded transfer functions must have a tail heavier than $x^{-1}$. Suppose that $\phi'(x) = O(x^{-p})$ as $x \to \infty$ for some $p > 1$. Then $\phi'(N^{1/2p}) = O(1/\sqrt{N})$ and so $(\phi')^{-1}(1/\sqrt{N}) = O(N^{1/2p})$ as $\phi'$ is asymptotically monotone decreasing. Therefore asymptotically

$$\mathbb{P}(\phi'(h)^2 S < 1/N) \sim ((\phi')^{-1}(1/\sqrt{N}))^{-\alpha} \tag{43}$$
$$\gtrsim N^{-\alpha/2p} \neq O(N^{\alpha/2-1}) \tag{44}$$

if $-\alpha/2p > \alpha/2 - 1$, i.e.

$$\alpha(1 + 1/p) < 2 \ . \tag{45}$$

This has a number of important implications. Firstly, *any sigmoidal transfer function is guaranteed to yield multifractal Jacobian eigenvectors in the stationary state for $\alpha \leq 1$*. We have seen above that binary transfer functions or those with a hard threshold already yield multifractality for all $\alpha < 2$. To determine the multifractality of Jacobian eigenvectors for $1 < \alpha < 2$ for the remaining sigmoidal transfer functions $\phi$, it suffices to check the asymptotic behaviour of $\phi'(x)$ as $x \to \infty$. For example, the figures in this Letter have used $\phi = \tanh$, for which $\phi'(x) = \mathrm{sech}(x)^2$. Since $\mathrm{sech}(x) = O(e^{-x})$, it follows that $\mathrm{sech}(x)^2 = O(x^{-p})$ for all $p > 1$, and so *the stationary Jacobian eigenvectors are multifractal for all $\alpha < 2$ under the transfer function $\phi = \tanh$*.

## III. ADDITIONAL RESULTS AND DETAILS

### A. Lévy dynamical mean-field theory

In the stationary state, a randomly selected local field

$$h_i \overset{i}{\sim} L(\alpha, 0, 0, g^\alpha \langle |\phi|^\alpha \rangle / 2) \tag{46}$$

is distributed as an $\alpha$-stable random variable whose scale parameter is computed self-consistently via

$$\langle |\phi|^\alpha \rangle = \int |\phi(h)|^\alpha p_{L(\alpha,0,0,g^\alpha \langle |\phi|^\alpha \rangle/2)}(h) \, dh \ . \tag{47}$$

This gives the distribution of $(\phi'(h_j))_j$ for use in the Jacobian cavity equations.

Taking the Fourier transform of the dynamic mean-field equation gives

$$i\omega h_i(\omega) = -h_i(\omega) + \sum_j J_{ij} F[\phi \circ h_j](\omega) \tag{48}$$

where $F$ is the Fourier operator and $h_i(\omega) := F[h_i](\omega)$. This complex equation may be broken into its real and imaginary parts to yield

$$\mathrm{Re}\, h_i(\omega) - \omega \,\mathrm{Im}\, h_i(\omega) = \sum_j J_{ij} \,\mathrm{Re}\, F[\phi \circ h_j](\omega) \ , \tag{49}$$

$$\mathrm{Im}\, h_i(\omega) + \omega \,\mathrm{Re}\, h_i(\omega) = \sum_j J_{ij} \,\mathrm{Im}\, F[\phi \circ h_j](\omega) \ . \tag{50}$$

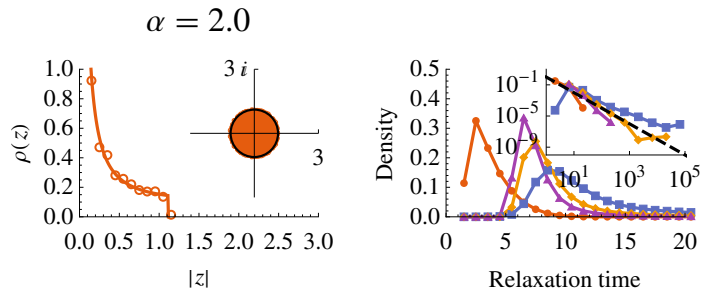

FIG. 3. Network statistics for classical homogeneous networks, cf. Fig. 2(a–b) of the main text.

Using the independence of the real and imaginary parts of the Fourier transform of $h_i(t)$ [12], along with the generalised central limit theorem, yields

$$D_\alpha(\mathrm{Re}\, h_i(\omega))_i + |\omega|^\alpha D_\alpha(\mathrm{Im}\, h_i(\omega))_i = \langle |\mathrm{Re}\, F[\phi \circ h_j](\omega)|^\alpha \rangle_j / 2 \,, \quad (51)$$

$$D_\alpha(\mathrm{Im}\, h_i(\omega))_i + |\omega|^\alpha D_\alpha(\mathrm{Re}\, h_i(\omega))_i = \langle |\mathrm{Im}\, F[\phi \circ h_j](\omega)|^\alpha \rangle_j / 2 \,, \quad (52)$$

where $D_\alpha$ is the scale parameter of the $\alpha$-stable random variable with the same asymptotic tail as the argument. When $\alpha = 2$, the operator $D_\alpha$ is reduced to become half the variance of its argument. Adding these two equations together and defining $\Delta(\tau) = 2F_\omega^{-1}[D_\alpha(\mathrm{Re}\, h_i(\omega))_i + D_\alpha(\mathrm{Im}\, h_i(\omega))_i](\tau)$ (which reduces to the regular autocorrelation $\Delta(\tau) = F_\omega^{-1}[\langle |h_i(\omega)|^2 \rangle_i](\tau) = \langle h_i(t) h_i(t+\tau) \rangle_{i,t}$ when $\alpha = 2$) gives

$$\Delta(\tau) + (-\Delta_\tau)^{\alpha/2} \Delta(\tau) = F_\omega^{-1}[\langle |\mathrm{Re}\, F[\phi \circ h_j](\omega)|^\alpha + |\mathrm{Im}\, F[\phi \circ h_j](\omega)|^\alpha \rangle_j](\tau) \,, \quad (53)$$

where $(-\Delta_\tau)^{\alpha/2}$ is the fractional Laplacian [13]. This reduces to the classical Newtonian equation [14] when $\alpha = 2$,

$$\Delta(\tau) - \frac{\partial^2}{\partial \tau^2} \Delta(\tau) = F_\omega^{-1}[\langle |F[\phi \circ h_j](\omega)|^2 \rangle_j](\tau) = \langle \phi(h_j(t)) \phi(h_j(t+\tau)) \rangle_{j,t} \,. \quad (54)$$

### B. The chaotic fluctuations of classical homogeneous random neural networks are delocalised

The localization of dominant Jacobian eigenmodes determines the spatial profile of network fluctuations and is described by the inverse participation ratio $\mathrm{IPR}_q(v) = \sum_i |v_i|^{2q}$ for a normalized vector $v$. Defining the multifractal dimension $D_q$ by the asymptotic relation $\mathrm{IPR}_q(v) \sim N^{(1-q)D_q}$ for large system size $N$, localized and delocalized states correspond to $D_q$ equaling zero and one respectively, while multifractal states are characterized by $D_q$ being a nontrivial function of $q$ (Fig. 1 of the main text), which are a hallmark of Anderson transitions [3]. Standard results in random matrix theory have established that the eigenvectors of random matrices with independent sub-exponential entries are delocalized [15], which includes the Jacobian matrices of classical homogeneous random neural networks ($\alpha = 2$). The chaotic fluctuations of classical random neural networks are thus delocalised, with the IPR (Fig. 2(d) of the main text, blue) staying close to the value $N^{1-q}$ attained for a constant, maximally delocalized vector $(1/\sqrt{N}, \ldots, 1/\sqrt{N})$, and the $D_q$ estimate staying close to that of a random $N$-dimensional spherical vector, which is delocalized with overwhelming probability [16] (Fig. 2(e) of the main text, blue and black); these two $D_q$ estimates remain slightly curved due to finite-size effects.

### C. Details of the reservoir computing task

The extended critical regime of correlated multifractal chaos has important computational implications for networks with heavy-tailed connectivity. To illustrate this, we consider a pattern classification task [17] with $P$ fixed patterns of randomly drawn binary vectors of length $L$. From these, $P$ pattern classes are formed by generating noisy realizations of each pattern with the addition of independent Gaussian vectors of variance $\sigma^2 = 0.1^2$. Each realization forms the initial state of the first $L$ neurons of the network with the connectivity and initial state of the remaining $N - L$ neurons remaining quenched for the task. At each timestep, a linear readout is trained for each pattern class by

linear regression to provide a unit output for the corresponding pattern and zero for the others, with the strongest readout signal determining classification. This task relies on the signal distance $H_{12}^{(s)}$, which is the Hamming distance between patterns of different classes, being greater than the noise distance $H_{12}^{(n)}$ between patterns of the same class. The difference between the signal and noise distances, $\Delta H_{12}$, is known as the linear separability of pattern classes, which is initially low [17]. Classical random neural networks are known to undergo transient chaotic dimensionality expansion (Fig. 3(b) of the main text, yellow) to increase class separability $\Delta H_{12}$ which aids computation in a reservoir computing context. However, the onset of chaotic mixing reduces separability $\Delta H_{12}$ and thus classification performance (Fig. 3 of the main text, bottom).

---



\end{document}